# Uncertainty in Island-based Ecosystem Services and Climate Change

Nazli Demirel[1]*, Ioannis N. Vogiatzakis[2,3], George Zittis[4], Mirela Tase[5], Attila D. Sandor[6,7,8], Savvas Zotos[2], Christos Zoumides[9], Turgay Dindaroglu[10], Mauro Fois[11], Irene Christoforidi[12], Valentini Stamatiadou[13], Shiri Zemah-Shamir[14], Tamer Albayrak[15], Cigdem Kaptan Ayhan[16], Paraskevi Manolaki[17], Ina Sieber[18], Ziv Zemah-Shamir[19], Elli Tzirkalli[2], Aristides Moustakas[20,21]

[1]Institute of Marine Sciences and Management, Istanbul University, 34820 Istanbul, Türkiye
[2]Faculty of Pure and Applied Sciences, Open University of Cyprus, Nicosia, Cyprus
[3]Department of Soil, Plant and Food Science, University of Bari Aldo Moro, Italy
[4]Climate and Atmosphere Research Center (CARE-C), The Cyprus Institute, Nicosia, Cyprus
[5]Department of Tourism, Aleksander Moisiu University, Durrës, Albania
[6]HUN-REN-UVMB Climate Change: New Blood-Sucking Parasites and Vector-Borne Pathogens Research Group, Budapest, Hungary
[7]Department of Parasitology and Zoology, University of Veterinary Medicine, Budapest, Hungary
[8]ISTAR-UBB Institute, Babes-Bolyai University, Cluj-Napoca, Romania
[9]Energy, Environment and Water Research Center (EEWRC), The Cyprus Institute, Nicosia, Cyprus
[10]Department of Forest Engineering, Karadeniz Technical University, Trabzon, Türkiye
[11]Department of Life and Environmental Sciences, University of Cagliari, Cagliari, Italy
[12]Department of Agriculture, Hellenic Mediterranean University, Heraklion, Greece
[13]Department of Marine Sciences, University of the Aegean, Mytilene, Greece
[14]School of Sustainability, Interdisciplinary Center (IDC), Reichman University, Herzliya, Israel
[15]Department of Biology, Dokuz Eylul University, İzmir, Türkiye
[16]Department of Landscape Architecture, Çanakkale Onsekiz Mart University, Çanakkale, Türkiye
[17]Deparment of Biology, Aarhus University, 8000 Aarhus C, Denmark
[18]Institute of Physical Geography and Landscape Ecology, Leibniz University, Hannover, Germany
[19]Department of Marine Biology, Leon H. Charney School of Marine Sciences, University of Haifa, Haifa, Israel
[20]Department of Product & Systems Design Engineering, University of the Aegean, Syros 84100, Greece
[21]Natural History Museum of Crete, University of Crete, Heraklion, Greece

*Corresponding author: ndemirel@istanbul.edu.tr

**Abstract**

Small and medium-sized islands are acutely exposed to climate change and ecosystem degradation, yet the extent to which uncertainty is systematically addressed in scientific assessments of their ecosystem services remains poorly understood. This study revisits 226 peer-reviewed articles drawn from two global systematic reviews on island ecosystem services and climate change, applying a structured post hoc analysis to evaluate how uncertainty is treated across methods, service categories, ecosystem realms, and decision contexts. Studies were classified according to whether uncertainty was explicitly analysed, just mentioned, or ignored. Only 30% of studies incorporated uncertainty explicitly, while more than half did not address it at all. Scenario-based approaches dominated uncertainty assessment, whereas probabilistic and ensemble-based frameworks remained limited. Cultural ecosystem services and extreme climate impacts exhibited the lowest levels of uncertainty integration, and few studies connected uncertainty treatment to policy-relevant decision frameworks. Weak or absent treatment of uncertainty emerges as a structural challenge in island systems, where narrow ecological thresholds, strong land–sea coupling, limited spatial buffers, and reduced institutional redundancy amplify the consequences of decision-making under incomplete knowledge. Systematic mapping of how uncertainty is framed, operationalised, or neglected reveals persistent methodological and conceptual gaps and informs concrete directions for strengthening uncertainty integration in future island-focused ecosystem service and climate assessments. Embedding uncertainty more robustly into modelling practices, participatory processes, and policy tools is essential for enhancing scientific credibility, governance relevance, and adaptive capacity in insular socio-ecological systems.

1. Introduction

Islands are increasingly recognized as frontline systems for understanding and responding to sustainability challenges. Their biophysical isolation, high levels of endemic biodiversity, strong socio-ecological dependencies, and exposure to compounding environmental risks make them unique laboratories for ecosystem service (ES) research (Balzan et al., 2018; Vogiatzakis et al., 2023). At the same time, islands face intensified pressures from climate change (CC), including sea-level rise, ocean acidification, and extreme weather events, alongside non-climatic drivers such as land-use change, tourism, pollution, and demographic shifts (Zittis et al., 2025; Moustakas et al., 2025). These overlapping challenges place exceptional demands on the robustness, usability, and policy relevance of ES assessments.

Over the past decade, ES frameworks have been increasingly applied to support sustainability strategies on islands, guiding conservation, adaptation, and development planning. This growing policy uptake, however, is often hindered by significant uncertainty. Key sources of uncertainty include ecological knowledge gaps, poor spatial and temporal resolution, limited stakeholder involvement, data scarcity, and methodological inconsistency (Hooftman et al., 2022). Although methodological advances such as ensemble modelling, probabilistic techniques, and scenario-based approaches offer powerful tools to address uncertainty, these remain underutilized in island-focused research (Baustert et al., 2018; Willcock et al., 2023). Given the ecological sensitivity and governance complexity of many island systems, the absence of robust uncertainty treatment can limit both the scientific and practical utility of ES assessments. These challenges are particularly pronounced in small and remote islands, where reduced monitoring infrastructure, limited accessibility to stakeholders, and a reliance on qualitative or perception-based data, especially for cultural and relational services, further complicate efforts to represent and manage uncertainty (IPCC, 2022; Walther et al., 2025). Moreover, the diversity of classification systems used in ES assessments, including frameworks such as the CICES (v5.1), MAES, and TEEB, can lead to additional inconsistency in how uncertainty is interpreted and reported (Haines-Young and Potschin-Young, 2018).

Recent systematic reviews have substantially advanced understanding of island-based ES assessments, particularly in terms of pressures, responses, and adaptation strategies under climate change (Zittis et al., 2025; Moustakas et al., 2025). While transparent screening protocols and consistent ES categorization approaches to synthesize empirical evidence across diverse island contexts were applied; they did not specifically evaluate how uncertainty itself is addressed within these assessments. This gap is significant, particularly as ES science seeks to improve its predictive capacity and decision relevance amid accelerating global change. Despite a growing body of ES and climate assessments, the systematic treatment of uncertainty in island contexts remains fragmented and under-theorized. This paper contributes to methodological innovation by offering a post-review classification of evidence gaps in how uncertainty is addressed across ES categories in island-focused studies. By re-examining an existing systematic evidence base, we move beyond descriptive synthesis to highlight recurring methodological blind spots, overlooked uncertainty dimensions, and weak links to policy translation. This approach not only identifies where current methods underperform but also lays the groundwork for more context-sensitive frameworks that can improve the credibility, transparency, and decision relevance of ES and climate assessments in islands.

This paper presents a structured post hoc analysis of 226 peer-reviewed articles drawn from the two aforementioned reviews. Our aim is to assess how uncertainty has been considered in empirical studies of ecosystem services and climate change on islands. Specifically, we examine (i) whether and how uncertainty is addressed methodologically; (ii) what types and sources of uncertainty are reported; and (iii) whether uncertainty is integrated into decision-making processes or intervention design. We also analyse variation in uncertainty treatment across ES categories, ecosystem types, climate drivers, and study designs. Through this analysis, we aim to identify structural patterns, highlight critical gaps, and propose methodological directions for advancing island-focused ES research.

## 2. Material and methods

This study builds upon two previously published systematic reviews (Zittis et al., 2025; Moustakas et al., 2025), which examined how island-based ES assessments address the pressures and responses linked to both climatic and non-climatic drivers. Here we performed a structured post hoc re-analysis of the empirical subset of these reviews, focusing specifically on how uncertainty is addressed within island-related ES and climate change research.

Islands are conceptualized as coupled socio-ecological systems characterized by physical isolation, bounded spatial extent, and strong land-sea connectivity. Operationally, we adopt the definition of islands established under the United Nations Convention on the Law of the Sea (UNCLOS, 1982), whereby islands are naturally formed land areas, surrounded by water, that remain above water at high tide and are capable of sustaining human habitation or economic life. Inland islands occurring within lakes or rivers are excluded, as are emergent rocks that do not support stable ecological or socio-economic systems. Beyond this legal definition, islands represent highly heterogeneous systems, varying substantially in size, remoteness, geomorphology, climatic exposure, biodiversity composition, and socio-economic structure, leading to pronounced diversity in ecosystem functioning, vulnerability, and adaptive capacity across regions. This heterogeneity is well documented in recent global syntheses of island ecosystem services and climate impacts, which highlight strong geographic, ecological, and socio-economic contrasts among island systems worldwide (Zittis et al., 2025; Moustakas et al., 2025). Accordingly, islands included in the analysis span a wide spectrum, from small low-lying coral atolls and volcanic oceanic islands to large continental islands and complex archipelagos.

The analysis focused on peer-reviewed scientific articles addressing ES and CC in island contexts. The initial literature search used the following Boolean operators: ("ecosystem service*" OR "ecosystem good*") AND (climate* NEAR chang*) AND (island* OR islet* OR archipelag*), applied as topic searches in Scopus and ISI Web of Science Core Collection. The final search was conducted in September 2024 and includes publications up to December 2023. The search was restricted to original research, letters, and reviews; specific ES terms (e.g. "crops", "fisheries") were not included to avoid excessive filtering. The PRISMA approach was used to eliminate duplicates, non-peer-reviewed and inaccessible items (Page et al., 2021; Haddaway et al., 2022). Studies that did not focus on islands, or where the term 'island' referred to another concept (e.g. "urban heat islands"), were excluded. This process resulted in 226 articles relevant to islands and ES under climate-related pressures (see Figure 1).

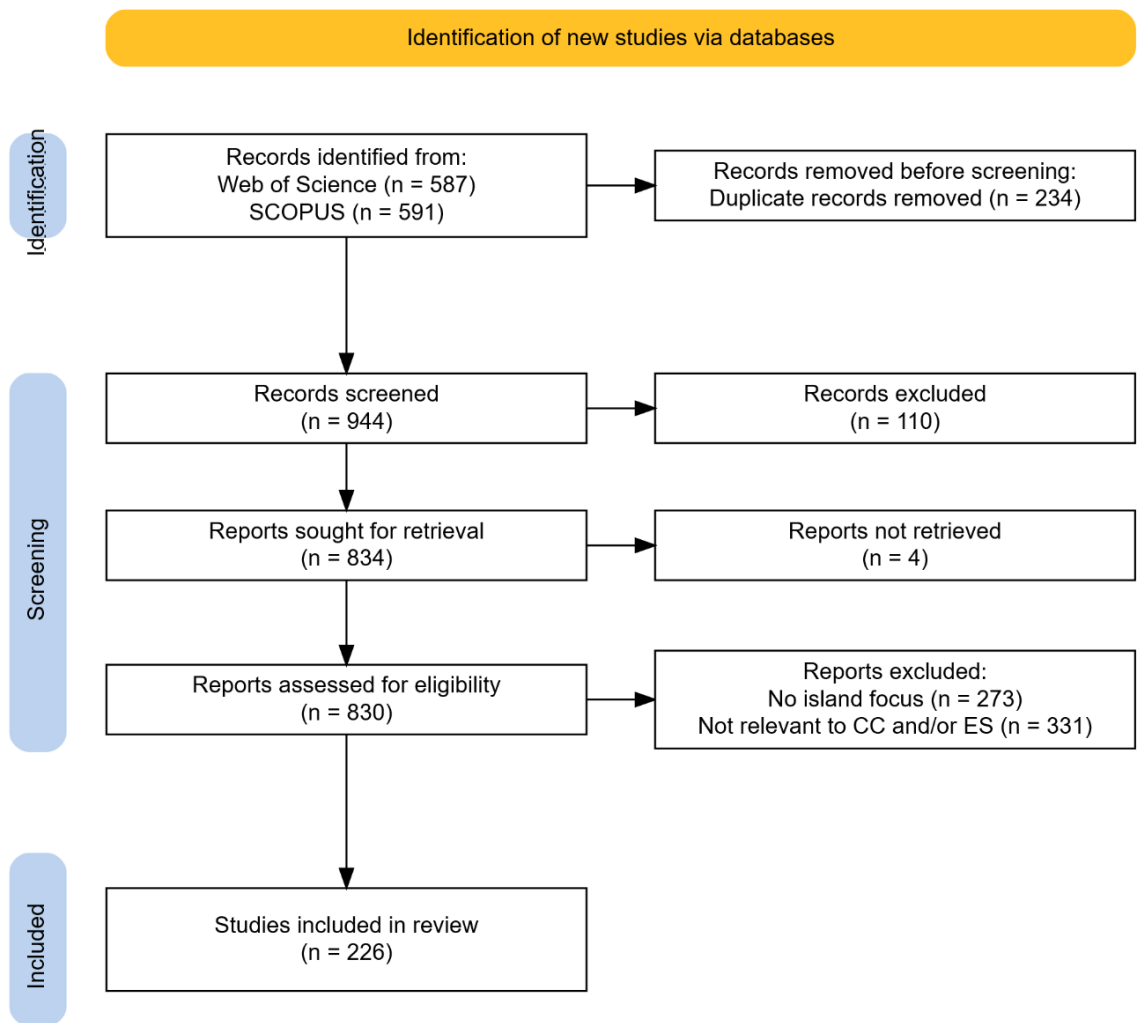

**Figure 1.** Workflow of the systematic literature review, following PRISMA method, forming the empirical basis for the post hoc uncertainty analysis.

Each article was then reviewed and classified based on its treatment of uncertainty, using a three-tiered typology (Table 1) adapted from recent uncertainty-focused syntheses that emphasize transparency and methodological rigor in ecosystem service research (Hamel and Bryant, 2017; Bryant et al., 2018; Walther et al., 2025). Classification was based on structured content analysis combining explicit terminology with interpretative assessment of methodological choices, analytical procedures, and reporting practices, rather than solely on the presence of the term "uncertainty". Studies were categorized as:

- Explicitly: Uncertainty was directly quantified, modelled, or systematically analyzed through probabilistic, sensitivity-based, ensemble, or scenario-based approaches;
- Mentioned: Uncertainty was acknowledged narratively or conceptually, without formal analytical treatment; and
- Ignored: No explicit reference to uncertainty was made and when study design and reporting did not allow inference of uncertainty treatment.

Descriptive analyses were then performed to examine how the treatment of uncertainty varied (Table 1) across: ES categories (provisioning, regulating, cultural, supporting); climate change impacts (e.g. sea-level rise, temperature increases, extreme events); ecosystem realms (terrestrial, marine, or coupled systems), and geographical locations according to world oceans including Mediterranean Sea. Figures were developed to visualize dominant methods, temporal trajectories, and thematic groupings.

**Table 1.** List of variables and values used in the analysis.

| Variables | Value |
|---|---|
| Uncertainty consideration | Explicitly<br>Just mentioned<br>Ignored |
| Study type | Climate Change impact<br>Ecosystem Service assessment |
| Ecosystem type | Terrestrial<br>Marine<br>Freshwater |
| Method of Uncertainty | Statistical<br>Scenarios<br>Multiple models |
| Ecosystem Services Assessment | ES: Food provision<br>ES: Raw materials<br>ES: Freshwater<br>ES: Medicinal resources<br>ES: Climate modulation<br>ES: Carbon sequestration<br>ES: Wastewater treatment<br>ES: Soil<br>ES: Pollination/Biological control<br>ES: Recreation/tourism<br>ES: Aesthetic appreciation<br>ES: Spiritual/Cultural |
| Climate Change impact | CC: Temperature<br>CC: Precipitation<br>CC: Sea level rise<br>CC: Ocean acidification<br>CC: Extremes |
|                    | CC: Other             |
| Decision making    | Considered            |
|                    | Mentioned             |
|                    | None                  |
| Impact Assessment  | Adaptive management   |
|                    | Cost/benefit analysis |

For the subset of studies that explicitly addressed uncertainty, additional thematic coding was undertaken. Methodological approaches were analyzed and categorized into three main groups: (i) scenario analysis, either exploratory or normative; (ii) multiple-model frameworks, including sensitivity testing or structural comparisons; and (iii) statistical techniques such as probabilistic estimations or the use of confidence intervals (Table 1). Attention was also given to whether uncertainty was linked to decision-making or intervention planning. Studies were accordingly classified as: explicitly linking uncertainty to decision support (e.g. through cost–benefit analysis or adaptive management); referencing uncertainty in the context of interventions without methodological integration; or not establishing any connection between uncertainty and interventions. Finally, the use of economic valuation or adaptive management frameworks to guide policy relevance was recorded, following established practices in the literature (Refsgaard et al., 2007; Walther et al., 2025).

To examine temporal patterns in how uncertainty is treated in island ecosystem service and climate change studies, we applied binary logistic regression models. Two binary outcomes were analyzed separately: (i) whether a study ignored uncertainty entirely, and (ii) whether a study explicitly incorporated uncertainty. Publication year was included as a continuous predictor and mean-centered to improve numerical stability. Two model specifications were fitted for each outcome. First, a year-only model was used to replicate a simple temporal trend analysis. Second, an adjusted multivariable model was fitted to account for major structural differences among studies. The adjusted models included study type (ecosystem service assessment only, climate change impact only, combined ES + CC, or neither), ecosystem realm (terrestrial, marine, freshwater; non-exclusive binary indicators), and broad geographic region (Pacific, Atlantic, Indian, Mediterranean, Polars, or Global), derived from the reported sea-zone of assessment.

Model performance was evaluated using likelihood-ratio (LR) $\chi^2$ tests comparing fitted models against the null (intercept-only) model, and by comparing Akaike Information Criterion (AIC) values between year-only and adjusted models. Results are presented as predicted probabilities with 95 % confidence intervals to facilitate interpretation of temporal trajectories. The analysis is descriptive in intent; publication year is treated as a temporal ordering variable rather than a causal driver.

3. Results

Out of the 226 reviewed studies, 68 (30%) explicitly considered uncertainty, 37 (16%) just mentioned it, and 121 (54%) ignored it (Figure 2A). In terms of ecosystem type, studies focusing on marine systems included the highest number of papers that explicitly considered uncertainty. Terrestrial studies had the highest number of papers categorized as ignoring uncertainty, while freshwater studies were more evenly distributed across the three categories (Figure 2B). Classification by study type showed that both CC impact studies and ES assessments included papers across all three uncertainty categories. A slightly higher number of ES assessments explicitly considered uncertainty, but in both groups, studies that ignored uncertainty had the largest share (Figure 2C). Among the 154 studies that assessed ES, food provision, raw materials, freshwater, and climate modulation were the most frequently associated with papers that explicitly considered uncertainty. Spiritual and cultural services, aesthetic appreciation, and recreation and tourism were mostly associated with studies that ignored or just mentioned uncertainty.

Pollination, soil-related services, and wastewater treatment showed a more mixed distribution across categories (Figure 2D). In the subset of CC impact studies (n = 154), temperature and precipitation were the most frequently associated with studies that explicitly considered uncertainty. Sea-level rise, ocean acidification, and extreme events were primarily linked to studies that either just mentioned or ignored uncertainty. The "other" category contained the largest number of studies classified as ignoring uncertainty (Figure 2E). When classified by ocean and sea regions, studies from the Pacific Ocean and Atlantic Ocean accounted for the largest number of papers across all three uncertainty categories. In both regions, the number of studies that ignored uncertainty exceeded those that explicitly considered it, particularly in the Pacific (49 ignored vs. 29 explicit) and Atlantic (28 ignored vs. 20 explicit). The Indian Ocean also showed a higher number of studies classified as ignoring uncertainty (10) compared to those explicitly considering it (6). In contrast, studies from the Mediterranean Sea exhibited a more balanced distribution across categories, while polar regions (Arctic and Antarctic) and global-scale studies were represented by comparatively small numbers of publications (Figure 2F).

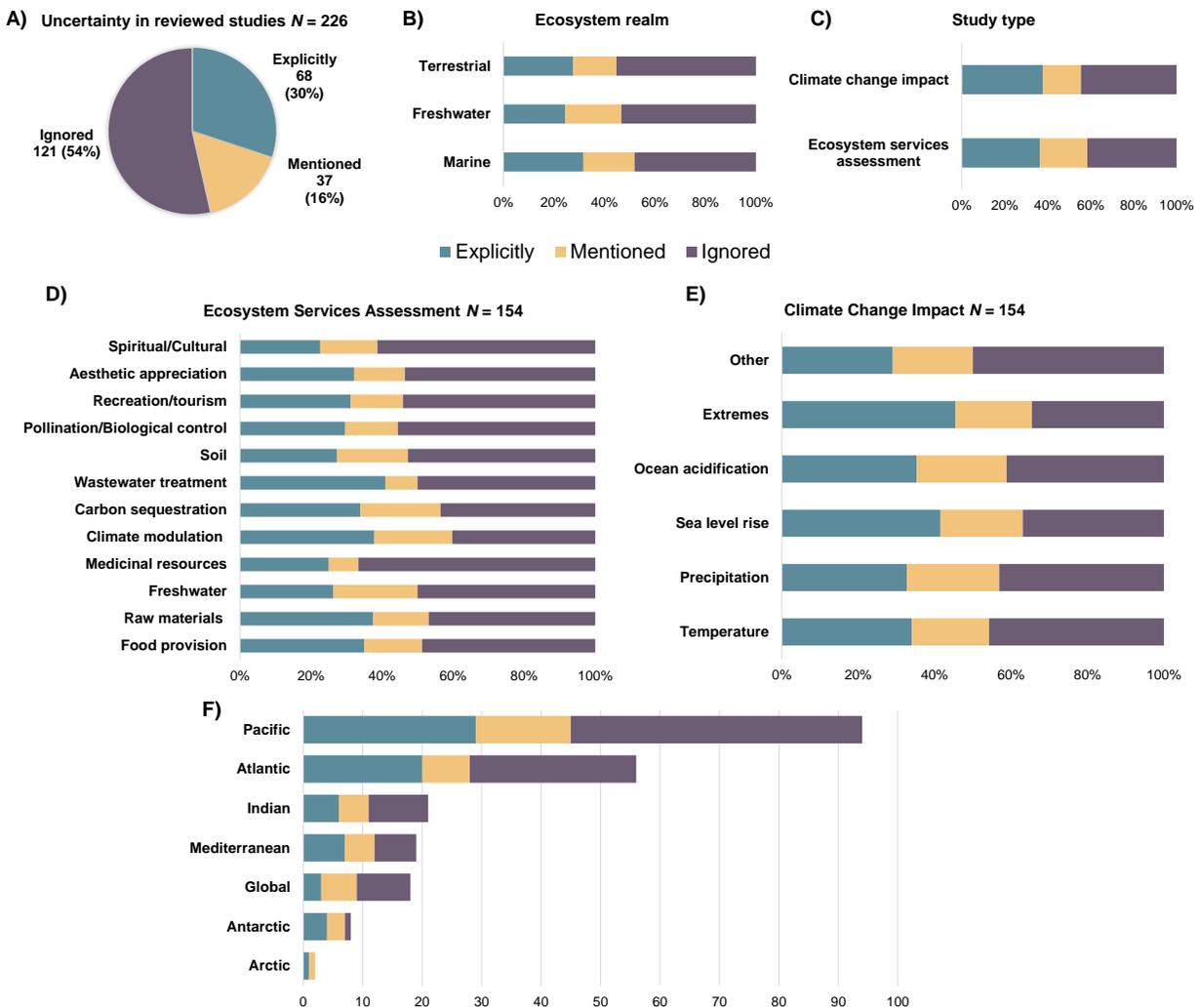

**Figure 2.** Classification of uncertainty consideration based on whether uncertainty was explicitly considered, just mentioned, or ignored: (A) in 226 reviewed papers, (B) by ecosystem type, (C) by study type, (D) across different ecosystem services domains (n = 154), and (E) across different climate change impact categories (n = 154), (F) location of assessment according to oceans including Mediterranean Sea.

Among the 68 studies that explicitly considered uncertainty, scenario-based approaches were the most frequently applied method (45%), followed by statistical methods (35%) and multiple model approaches (20%) (Figure 3A). In studies that also addressed decisions or interventions (n = 54), scenario approaches were most commonly used regardless of whether decision-making was considered, mentioned, or not addressed. Statistical methods were more prevalent when decisions and interventions were not discussed, while multiple models were used less frequently across all categories (Figure 3B). Among the 46 studies that assessed the impact of interventions, scenario-based methods were most commonly applied in studies using adaptive management. In contrast, statistical approaches were more frequently used in cost–benefit analyses. Multiple models were employed in both contexts but remained the least used method overall (Figure 3C). Across the 56 ES assessments, scenario-based methods were commonly applied in studies focusing on provisioning and regulating services such as food provision, raw materials, freshwater, climate modulation, and carbon sequestration. Statistical methods were more prominent in assessments of cultural services, pollination and biological control, soil, and wastewater treatment. Multiple models were used across a range of ES domains but at lower frequencies (Figure 3D). In CC impact studies (n = 62), scenarios were most frequently applied in the assessment of temperature, precipitation, and sea-level rise. Statistical methods were more commonly used in relation to ocean acidification, extreme events, and other categories. The use of multiple models was present across all categories but remained limited (Figure 3E).

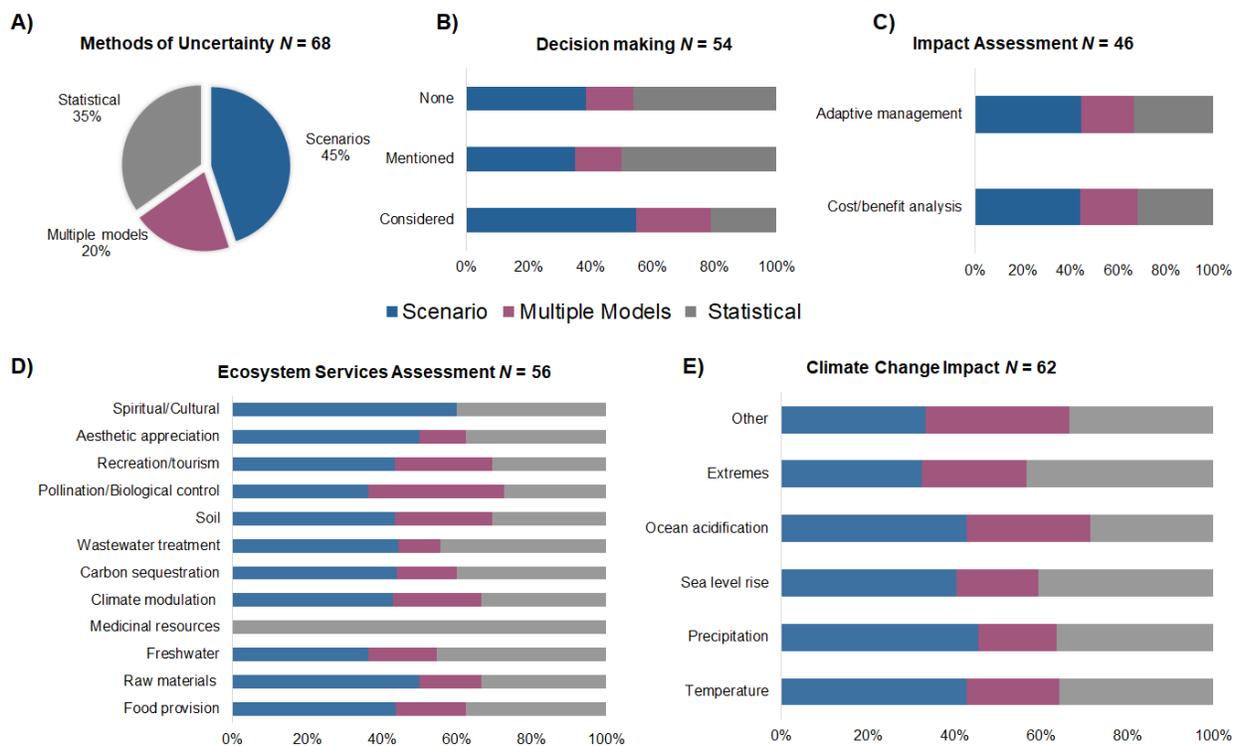

**Figure 3.** Distribution of uncertainty assessment methods in studies that explicitly considered uncertainty: (A) overall methods used (n = 68), (B) by whether decisions and interventions were considered, mentioned, or not addressed (n = 54), (C) by type of impact assessment method (n = 46), (D) across different ecosystem services (n = 56), and (E) across different climate change impact categories (n = 62).

The probability that studies ignored uncertainty showed a modest decline over time (Figure 4a). In the year-only model, inclusion of publication year resulted in a small improvement in model fit relative to the null model (LR $\chi^2$ = 3.04, p = 0.081), indicating a weak downward temporal tendency. However, after accounting for study type, ecosystem realm, and geographic region, the temporal effect weakened further and remained non-significant (LR $\chi^2$ = 12.76, p = 0.078). Comparison of information criteria indicated no improvement in overall model fit when moving from the year-only to the adjusted model (AIC = 309.1 vs. 311.4), suggesting that variation in uncertainty omission is largely explained by study characteristics rather than publication year alone. In contrast, no temporal trend was detected for the explicit incorporation of uncertainty (Figure 4b). The year-only model did not differ from the null model (LR $\chi^2$ = 0.07, p = 0.798), indicating no evidence of increasing explicit uncertainty treatment over time. In the adjusted model, overall model fit improved relative to the null (LR $\chi^2$ = 14.72, p = 0.040), reflecting the influence of study type, ecosystem realm, and region. However, publication year itself remained non-significant, and the adjusted model showed only a marginal improvement in information criteria compared to the year-only model (AIC = 277.7 vs. 280.4).

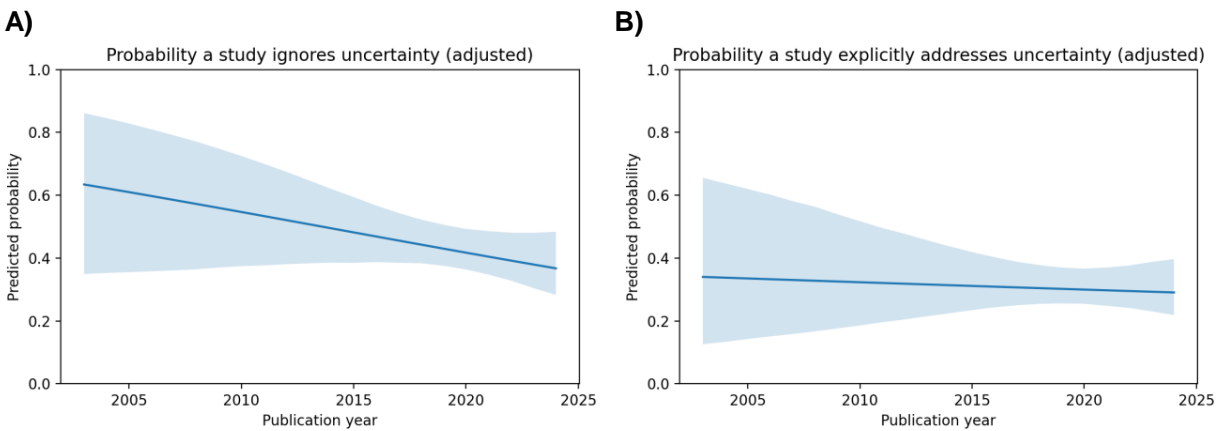

**Figure 4.** Temporal trends in the treatment of uncertainty in island ecosystem service and climate change studies based on adjusted binary logistic regression. (A) Predicted probability that studies ignore uncertainty, and (B) predicted probability that studies explicitly incorporate uncertainty, plotted against publication year. Predictions are derived from multivariable models that include publication year while accounting for differences in study type, ecosystem realm, and geographic region. Solid lines represent model-based predicted probabilities, and shaded bands indicate 95% confidence intervals. Publication year is treated as a temporal ordering variable rather than a causal driver.

4. Discussion

The uneven treatment of uncertainty across island-focused ecosystem services and climate change studies reveals a persistent tension between scientific modelling and practical decision-making in data-limited, ecologically distinct contexts. Although methodological advances, such as scenario planning, ensemble modelling, and probabilistic techniques, are increasingly available, fewer than half of the reviewed studies provided a meaningful methodological treatment of uncertainty. This reinforces concerns expressed in the literature that assessments omitting uncertainty, or addressing it only superficially, risk diminished credibility and reduced policy relevance, particularly in highly sensitive and rapidly changing geographies (Refsgaard et al., 2007; Cash et al., 2003; Walther et al., 2025). While scenario-based approaches were the most frequently applied, their use was often restricted to exploratory foresight, with limited incorporation of probabilistic modelling, model comparison, or sensitivity testing. These findings align with earlier reviews of ecosystem services modelling (Runting et al., 2017; Baustert et al., 2018), and continue to raise concerns regarding the robustness and interpretability of assessments that are intended to inform sustainability transitions in small island contexts. The temporal analysis further indicates that these patterns are not strongly driven by publication year alone. After accounting for differences in study type, ecosystem realm, and geographic region, the declining tendency for studies to ignore uncertainty weakened, and no systematic increase in the explicit methodological treatment of uncertainty was detected. This suggests that while awareness of uncertainty may be increasing rhetorically, its operational integration remains uneven and strongly conditioned by the structural characteristics and focus of studies, rather than by temporal progression per se.

Island systems face distinct and often compounding sources of uncertainty. These include narrow ecological baselines, steep land–sea gradients, heightened exposure to climate-driven impacts, and lower institutional redundancy. Such conditions amplify the consequences of downplaying or omitting uncertainty in both scientific analyses and policy decisions. Despite this, only a minority of studies acknowledged the potential limitations of applying global models or continental assumptions to insular settings, where ecological and social thresholds may be crossed more rapidly and with reduced adaptive capacity (Snell et al., 2018; Moustakas et al., 2025). While some studies incorporated climate-sensitive variables such as sea-level rise, precipitation variability, or ocean acidification, few reported confidence intervals or explored how uncertainty propagates through ecosystem service estimates. Moreover, the limited integration of extreme event dynamics, despite their increasing influence on coastal and marine systems, underscores a continued gap in aligning risk assessment practices with the spatial and temporal realities of island governance (Zittis et al., 2025; Hernández-Delgado, 2024). Taken together, these characteristics mean that decision-making under unacknowledged or weakly characterized uncertainty can generate disproportionately large ecological and societal consequences in island systems compared to continental contexts. Limited spatial buffers restrict opportunities for risk displacement, narrow ecological thresholds increase the likelihood of rapid regime shifts, and strong land-sea coupling facilitates the propagation of disturbances across ecosystem compartments. At the same time, reduced institutional redundancy and governance capacity constrain adaptive responses, amplifying the costs of maladaptation. In this sense, uncertainty in islands does not merely represent a technical modelling challenge but constitutes a core structural dimension of vulnerability, directly shaping resilience, sustainability trajectories, and policy risk.

A differential pattern emerged across ES categories in how uncertainty was framed in islands. Provisioning and regulating services, particularly those related to fisheries, freshwater availability, and carbon storage, were more frequently accompanied by explicit uncertainty assessments, most often through scenario modelling or multi-baseline analyses. In contrast, cultural services and less tangible regulating services (e.g. spiritual value, sense of place, landscape aesthetics) were rarely approached with any structured uncertainty framework. This pattern reflects a broader bias within ecosystem service assessments, where non-material or relational values are often underrepresented, and this underrepresentation is further exacerbated when uncertainty is excluded altogether. In island contexts, where cultural heritage, customary marine tenure, and spiritual dimensions are integral to how ecosystems are valued and used, this absence of uncertainty framing can lead to the marginalization of local priorities in management and spatial planning processes (Solé Figueras et al., 2024; Bennett et al., 2024).

Island characteristics are expected to shape both ecosystem service composition and the way uncertainty is produced and communicated. Differences in island size, remoteness, exposure regimes, and socio-economic conditions strongly influence which ecosystem services dominate livelihoods, governance priorities, and vulnerability profiles, while also constraining data availability and the feasibility of fine-scale modelling (Balzan et al., 2018; Vogiatzakis et al., 2023). Scale-related issues are particularly difficult to address because studies differ widely in spatial coverage, ranging from partial land or marine areas to entire islands, while surface areas vary by several orders of magnitude across case studies, limiting comparability and generalization (Snell et al., 2018). Smaller and more remote islands frequently rely on coarse-resolution climate and ecosystem datasets that fail to capture localized variability, and logistical constraints associated with remoteness contribute to persistent geographic research gaps (Moustakas et al., 2025; Zittis et al., 2025). In our post hoc re-analysis, variability linked to island characteristics is therefore only indirectly reflected through descriptors that could be extracted consistently across studies, including ecosystem realm (terrestrial, marine, freshwater), ecosystem service category, climate driver type, and regional aggregation by sea zone.

While stakeholder perspectives and participatory approaches were frequently mentioned across the dataset, their actual integration into uncertainty assessment was limited. There was a notable absence of co-produced modelling frameworks, iterative planning processes, or structured deliberation around the limits of data and models. This is particularly problematic in island settings, where lived experience, traditional knowledge, and community-based governance systems play a central role in environmental stewardship (Balzan et al., 2018; Kenter et al., 2011). Without mechanisms to articulate preference uncertainty, institutional constraints, or value-based trade-offs, scenario exercises risk becoming detached from the realities of decision-making. This is consistent with emerging evidence that stakeholder perceptions of drivers and impacts, including distinctions between land-use and climate change, can influence the framing of ecosystem service outcomes (Moustakas et al. 2026). Participatory methods for uncertainty evaluation, such as confidence mapping, multi-criteria analysis, or stakeholder-elicited parameter testing, remain rarely applied, despite their potential to enhance relevance and trust in modelling outputs (Latour & van Laerhoven, 2024; Galaitsi et al., 2024).

Operationalizing uncertainty in islands within policy-relevant tools also remains weak across the reviewed literature. Although some references were made to adaptive management frameworks, very few studies included formal mechanisms to evaluate trade-offs, robustness, or opportunity costs under uncertainty. An even smaller subset linked uncertainty to actionable decisions such as defining protection thresholds, prioritizing restoration options, or assessing infrastructure investment risks in islands. This is particularly concerning in the context of small islands, where spatial constraints, path-dependency, and limited institutional capacity make policy missteps harder to reverse (Watson et al., 2021; Minderhoud et al., 2025). Furthermore, global thresholds embedded in sustainability models, such as 2°C temperature rise or 0.5 m sea-level increase, can produce qualitatively different implications for island systems. Even relatively small uncertainties in such projections can shift the balance between adaptation, relocation, or inaction, with lasting implications for ecosystem and human security (Katz et al., 2013; Benavides Rios et al., 2024).

This review complements recent island-specific systematic syntheses by Zittis et al. (2025) and Moustakas et al. (2025), which examined uncertainty in climate and ecosystem modelling across wider geographies. Our contribution, on the other hand, lies in systematically unpacking how uncertainty is addressed or neglected across ecosystem service types within island contexts, highlighting patterns that remain obscured in broader-scale assessments. By categorising methods, themes, and policy linkages, this review identifies persistent gaps: notably, the lack of uncertainty framing in cultural services, the underuse of participatory approaches, and the methodological opacity of scenario-based valuation studies. Still, several limitations must be acknowledged. This review focused solely on peer-reviewed literature in English, likely underrepresenting place-based knowledge, grey literature, or insights published in regional languages. Moreover, this remains a qualitative synthesis. While thematic patterns are consistent and meaningful, no claims are made about quantitative effect sizes or statistical generalisability. Importantly, our review is shaped by the scope of the original systematic reviews from which the dataset was drawn (Zittis et al., 2025; Moustakas et al., 2025). Those syntheses clearly demonstrated that peer-reviewed island research remains geographically uneven, with a strong concentration in a limited number of regions, particularly across the Pacific and selected Atlantic basins, while many small, remote, and developing island states remain weakly represented. This structural asymmetry reflects broader disparities in research infrastructure, funding access, and publication capacity, rather than the absence of climate-related impacts or ecosystem transformations. Furthermore, these source studies did not systematically trace how uncertainty propagates through modelling components, nor did they distinguish between epistemic and aleatory forms of uncertainty, limitations that also constrain our re-analysis. As such, dimensions such as input data quality, scenario consistency, or model structural validity fall outside the scope of this synthesis. Future work could address these gaps by incorporating multilingual searches, grey literature reviews, and structured expert elicitation to better reflect locally embedded knowledge and practitioner-led innovations.

Overall, our findings underscore the scientific and ethical importance of treating uncertainty as a central feature of climate and ecosystem service assessments in island contexts. The fragmented and inconsistent integration of uncertainty, across methods, service types, and geographies, limits the interpretability and practical utility of these studies. This review advances the field by mapping how uncertainty is framed, quantified, and operationalised across island assessments, offering a foundation for more context-sensitive, participatory, and robust evaluation frameworks. For small islands, characterised by unique socio-ecological coupling and heightened exposure to global change, embedding uncertainty into assessment design, scenario modelling, and stakeholder engagement is not merely a methodological imperative, it is essential to building just, adaptive, and durable sustainability pathways.

## 5. Conclusions and Recommendations

For islands, addressing uncertainty is not simply a methodological refinement but a prerequisite for credible, just, and adaptive sustainability pathways. This review demonstrates that while uncertainty is widely acknowledged in theory, its operationalization across ES categories and decision frameworks remains uneven. To enhance robustness and policy relevance, we recommend some further steps for future research and practice:

1. Island-specific assessment frameworks should be developed by adapting ensemble and probabilistic approaches to contexts where data scarcity and quality constraints prevail, including the use of simplified or hybrid modelling techniques tailored to insular conditions.
2. Participatory mechanisms must be strengthened through the systematic incorporation of local and traditional knowledge into uncertainty assessments, using co-produced models, participatory scenario development, and structured deliberation on data and model limitations.
3. The treatment of cultural ecosystem services should be expanded by embedding uncertainty framing into assessments of relational and non-material values, acknowledging their essential role in island livelihoods, identity, and governance systems.
4. Uncertainty should be operationalized more directly within policy tools, linking it explicitly to decisions around spatial zoning, restoration prioritization, and climate adaptation strategies, to ensure that assessments inform actionable outcomes.
5. Island typologies, harmonized spatial metadata, and downscaled climate and ecosystem service datasets, designed specifically for insular contexts, particularly for small and remote islands, should be integrated into future climate change and ecosystem service assessments.
6. Investment in regional data infrastructure and capacity building is needed, including the development of collaborative data hubs, open-access repositories, and training programmes that enhance the ability of island institutions to manage, interpret, and communicate uncertainty effectively.

ES and climate assessments in island contexts needs to be more transparent, credible, and directly useful to policymakers and communities. Ultimately, integrating uncertainty is not only a technical necessity but also an ethical imperative, ensuring that decisions made under conditions of risk and constraint support the long-term resilience of both ecosystems and societies.

**Acknowledgements**

This research was supported by the COST Action SMILES (CA21158): Enhancing Small-Medium IsLands resilience by securing the sustainability of Ecosystem Services.